\documentstyle[12pt]{article}
\begin{document}

\centerline{\bf Self-Attracting Walk on Lattices }
\centerline{$\;$}
\centerline{Jae Woo Lee\footnote{e-mail:jwlee@munhak.inha.ac.kr}}
\centerline{$\;$}
\noindent{Department of Physics, Inha University,
Inchon 402-751, Korea}
\centerline{$\;$}
\noindent{Department of Physics, Clarkson University, 
Potsdam, NY 13699-5820, USA}

\begin{abstract}
We have studied a model of self-attracting walk proposed by Sapozhnikov
using Monte Carlo method. The mean square displacement $ \langle R^2(t) \rangle \sim
t^{2\nu}$ and the mean number of visited sites $ \langle S(t) \rangle  \sim t^{k}$ are
calculated for one-, two- and three-dimensional lattice. In one dimension,
the walk shows diffusive behaviour with $\nu=k=1/2$. However, in two
and three dimension, we observed a non-universal behaviour, i.e., the
exponent $\nu$ varies continuously with the strength of the attracting
interaction.\\
\end{abstract}

\noindent{PACS:\ \ 05.40.+j, 02.50.-r, 05.60.+w}

\newpage

There is great interest in random walks and interacting
walks[1,2,3]. 
The mean square displacement of a random walk, $ \langle R^2(t)
\rangle $ follows a power-law $ \langle R^2(t) \rangle  \sim
t^{2\nu} $.
The ordinary random walk(RW) is diffusive, with
$\nu=1/2$, in all dimensions. For a walk with repulsion, such as
self-avoiding walk(SAW), the exponent $\nu$ is greater than 1/2. 
Random walk on a fractal is anomalous with $\nu < 1/2$[2,3].
Various models of interacting walks have been studied, such as true
self-avoiding walk[4,5], generalized true SAW[6],
the Domb-Joyce model[7], and an interacting walk with a weight factor
$p$ for each new site that the random walker visits[8,9].
A comparative study of interacting random walk models was
performed by Duxbury et.al.[10,11].

Recently, Sapozhikov proposed a generalized walk in which the
probability for the walker to jump to a given site is proportional to
$p=\exp(-n u)$, where $n=1$ for the sites visited by the walker at least
once and $n=0$ for other sites[12]. If $u< 0$, the walker is attracted to its
own trajectory. This walk is called a self-attracting walk(SATW). Monte
Carlo studies have suggested that $\nu < 1/2$ for $u=-1$ and $u=-2$ on two
dimension and $1/4 < \nu < 1/3$ on three dimension[12].
However, Aar\~{a}o Reis obtained non-universal behaviour of the SATW in
one to four dimension using exact enumeration method[13]. Prasad
et.al. concluded that the SATW in one dimension is diffusive[14].

In the present comment we report results of a  Monte Carlo simulation
for the self-attracting walk in one,
two and three dimension. We find that SATW in one dimension is
diffusive. However, SATW in two and three dimension shows
non-universal behaviour. Monte Carlo simulations were performed on one
dimensional lattice ($10^6$-steps with 2000-configurational averages for each
parameter $u$), square lattice ($10^6$-steps with 2000-averages), and
cubic lattice ($10^5$-steps with 2000-averages). The lattice sizes
used in this simulation were $L=10^6$ (1D), $1024 \times 1024$ (2D) and
$200 \times 200 \times 200$(3D). We always used the periodic boundary
conditions. We calculated the mean square displacement $ \langle R^2(t) \rangle$ and
the mean number of visited sites $ \langle S(t) \rangle $. 

Fig. 1 (a) shows the log-log plot of the mean square displacement
against the time. We obtained the exponent by a least-square fit:
$\nu=0.500(7)$ for $u=0$, $\nu=0.500(9)$ for $u=-0.5$, $\nu=0.499(8)$
for $u=-1.0$, and $\nu=0.498(9)$ for $u=-2.0$. All the lines are
parallel in the large-time limit. These results means that SATW in one
dimension is diffusive and support the conclusions of Prasad
et.al.[14]. This diffusive behaviour is further supported by the
results of the mean number of visited sites,
 $ \langle S(t) \rangle $, in Fig. 1 (b).
The mean number of visited sites $ \langle S(t) \rangle $ 
shows the power law behaviour as $ \langle S(t) \rangle  \sim t^{k}$
with
$k=0.499(7)$ for $u=0.0$, $u=0.499(4)$ for $u=-0.5$, $k=0.499(8)$ for
$u=-1.0$, and $k=0.498(5)$ for $u=-2.0$. At short times the slopes of
the mean square displacement depend on the parameter $u$. However, for
large times all the lines are parallel and give the same slope
regardless of the parameter $u$. Aar\~{a}o Reis concluded that the
exponent $\nu$ in one dimension decreases continuously when $u$
decreases, by the exact enumeration calculation up to $t=30$[13]. But
their conclusion is not correct because the time steps are too short to
reach the asymptotic behaviour,
 and the log-log plot of $ \langle R^2(t) \rangle$ has curvature
 at short times.

The log-log plot of the mean-square displacement and the mean number of
visited sites of a two dimensional SATW versus time
 are shown in Fig. 2 (a) and (b), respectively.
The appearance of the plateau region for large times
 is due to the finite
size of the substrate. In this region the walk touches the boundary.
The exponents $\nu$ were
obtained as $\nu=0.500(3)$ for $u=0$, $\nu=0.472(4)$ for $u=-0.5$,
$\nu=0.404(5)$ for $u=-1.0$, and $\nu=0.300(9)$ for $u=-2.0$.
Our results are consistent with those of Aar\~{a}o Reis[13],
Sapozhnikov[12], and Lee[15]. Values of the exponents $\nu$ decrease
continuously when $u$ decreases. We can not observe a critical value
$u_c$ proposed Sapozhnikov so that $\nu =1/2$ for $0 < u < u_c$[12].
The mean number of visited sites $ \langle S(t) \rangle $ shows the logarithmic
correction for random walk ($u=0$ case)[1,2]. 
The slopes of the log-log plots of $ \langle S(t) \rangle $
decrease continuously when $u$ decreases. 

Fig. 3 (a) and (b) show a log-log plot for the mean square displacement
and the mean number of visited sites for a three dimensional SATW in
a cubic lattice. 
The plateau region is also due to the finite size of the substrate.
The exponents $\nu$ were obtained as $\nu=0.500(2)$ for
$u=0$, $\nu=0.485(4)$ for $u=-0.5$, $\nu=0.466(5)$ for $u=-1.0$, and
$\nu=0.294(6)$ for $u=-2.0$. The exponents $\nu$ also decrease
continuously when $u$ decreases. These results are also consistent with
those of Aar\~{a}o Reis[13]. But they are not consistent with the prediction
of Sapozhnikov that $1/4 < \nu < 1/3$. 
In Sapozhnikov's argument there is an ambiguity regarding the scaling between
the bulk cluster and the boundary cluster visited by the walk. 
We obtained the
exponents $k$ as $k=0.996(5)$ for $u=0$, $k=0.993(2)$ for $u=-0.5$,
$k=0.991(5)$ for $u=-1.0$, and $k=0.904(6)$ for $u=-2.0$.
For random walk, $k=1$ in three dimension[1,2]. 
We observed that the values of $k$ decrease slowly when $u$ decreases.
These Monte Carlo results are also in good agreement with the
observations of Aar\~{a}o Reis[13].
The continuous decrease of the exponent $\nu$ and $k$ in 
three dimension is further 
confirmed by the simulation at the $u$-value between $u=-1.0$ and
$u=-3.0$. We obtained the exponents as $\nu=0.436(4)$, $k=0.970(3)$ for
$u=-1.5$, $\nu=0.240(5)$, $k=0.806(3)$ for $u=-2.5$, and
$\nu=0.190(8)$, $k=0.776(5)$ for $u=-3.0$. These results support that
there are no crossover behaviours at the range
$ -3.0 \le u \le 0 $. 

We calculated the exponents $\nu$ and $k$ in one, two and three
dimensions by Monte Carlo simulations. We have concluded that SATW in one
dimension is diffusive. This observation supports the recent calculation
of Prasad et.al.[14]. However, in two and three dimensions the
exponents $\nu$ decrease continuously when $u$ decreases. These
non-universal behaviours are consistent with those of Aar\~{a}o Reis[13].
Our observations are limited on regular lattices. One still has to explore
the scaling behaviour on general lattices such as fractal.

{\sl This work has been supported  by Inha University and by the Basic
Science Institute Program, Ministry of Education, Project No.
BSRI-97-2430.
I wish to thank professor Vladimir Privman for his careful reading of
this manuscript and good comments.}

\newpage
\centerline{\bf References}

\ 

\frenchspacing{

\noindent\hang [1] J.Haus and K.W. Kehr, Phys. Rep. {\bf 150},
2631 (1987).

\noindent\hang [2] S. Havlin and D. ben-Avraham, Adv. Phys. {\bf 36},
695 (1987).

\noindent\hang [3] A. Bunde and S. Havlin, ``Fractal and Disordered
Systems'', (Springer-Verlag, Berlin 1991).

\noindent\hang [4] D.J. Amit, G. Parisi and  L. Pelit, Phys. Rev. B
{\bf 27}, 1635 (1983).

\noindent\hang [5] A.L. Stella, S.L.A. de Queiroz, P.M. Duxbury and
R.B. Stinchcome, J. Phys. A:Math. Gen. {\bf 17}, 1903 (1984).

\noindent\hang [6] H.C. \~{O}ttinger, J. Phys. A:Math. Gen. {\bf 18},
L363 (1983).

\noindent\hang [7] C. Domb and G.S. Joyce, J. Phys. C:Solid State Phys.
{\bf 5}, 956 (1972).

\noindent\hang [8] H.E. Stanley, K. Kang, S. Redner and R.L. Blumberg,
Phys. Rev. Lett. {\bf 51}, 1223 (1983).

\noindent\hang [9] S. Redner and K. Kang, Phys. Rev. Lett. {\bf 51},
7129 (1983).

\noindent\hang [10] P.M. Duxbury, S.L.A. de Queiroz and R.B.
Stinchcome, J. Phys. A:Math. Gen. {\bf17}, 2113(1984).

\noindent\hang [11] P.M. Duxbury and S.L.A. de Queiroz, J. Phys.
A:Math. Gen. {\bf 18}, 661 (1985).

\noindent\hang [12] V.B. Sapozhnikov, J. Phys. A:Math. Gen. {\bf 27},
L151(1994).

\noindent\hang [13] F.D.A. Aar\~{a}o Reis, J. Phys. A:Math. Gen. {\bf
28}, 3851 (1995).

\noindent\hang [14]  M.A. Prasad, D.P. Bhatia and D. Arora, J. Phys.
A:Math. Gen. {\bf 28}, 3037 (1996).

\noindent\hang [15] J.W. Lee, J. Kor. Phys. Soc. {\bf 28}, S403 (1995).

}
\newpage
\centerline{\bf Figure Captions}

\ 

\noindent\hang Figure 1:
(a) Log-log plot of the mean square displacement versus time;
(b) log-log plot of the mean number of visited sites versus time
for one dimensional SATW
for $u=0$ (topmost curve), $-0.5$, $-1.0$, $-2.0$ (bottom
curve).

\ 

\noindent\hang Figure 2:
(a) Log-log plot of the mean square displacement versus time;
(b) log-log plot of the mean number of visited sites versus time
for two dimensional SATW
for $u=0$ (topmost curve), $-0.5$, $-1.0$, $-2.0$ (bottom
curve).

\ 

\noindent\hang Figure 3:
(a) Log-log plot of the mean square displacement versus time;
(b) log-log plot of the mean number of visited sites versus time
for three dimensional SATW
for $u=0$ (topmost curve), $-0.5$, $-1.0$, $-2.0$ (bottom
curve).

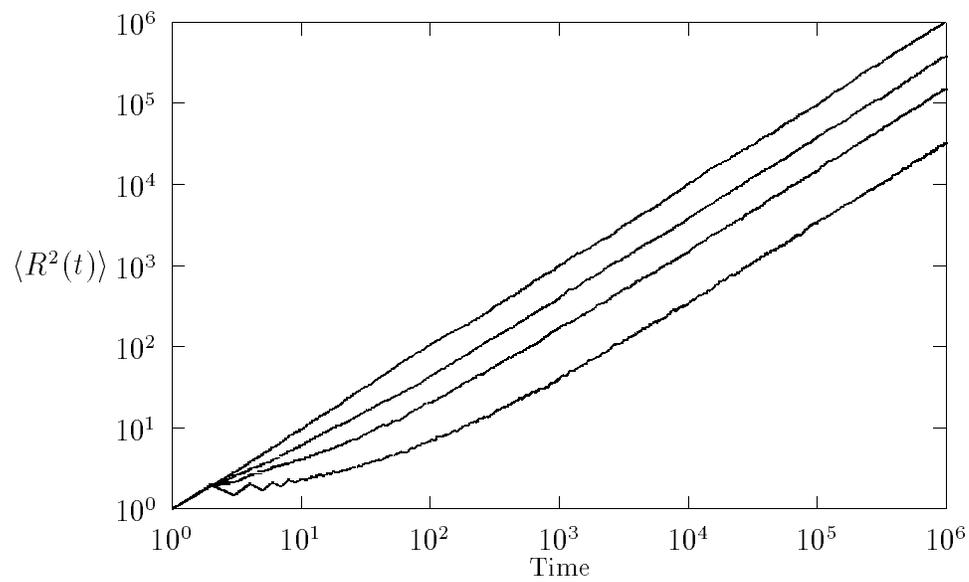
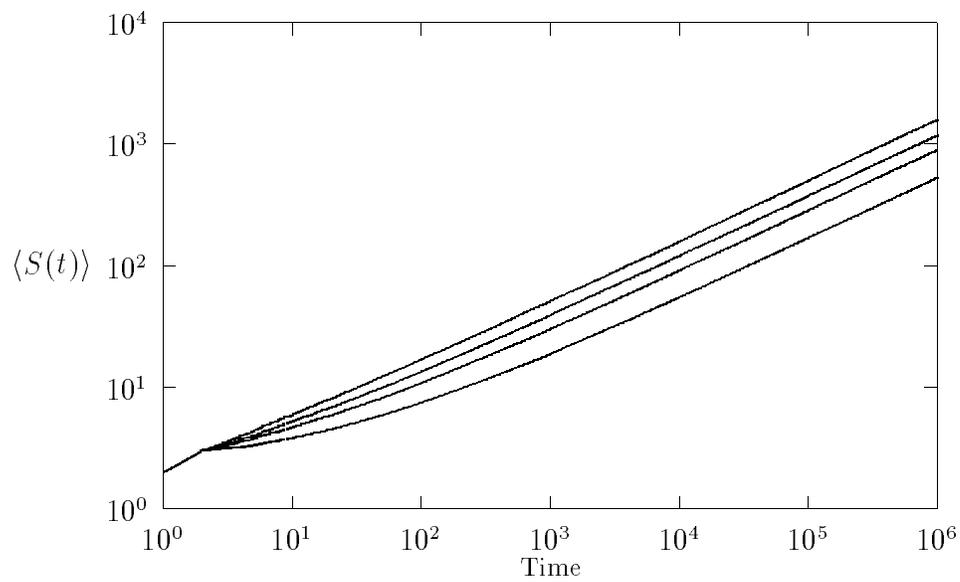
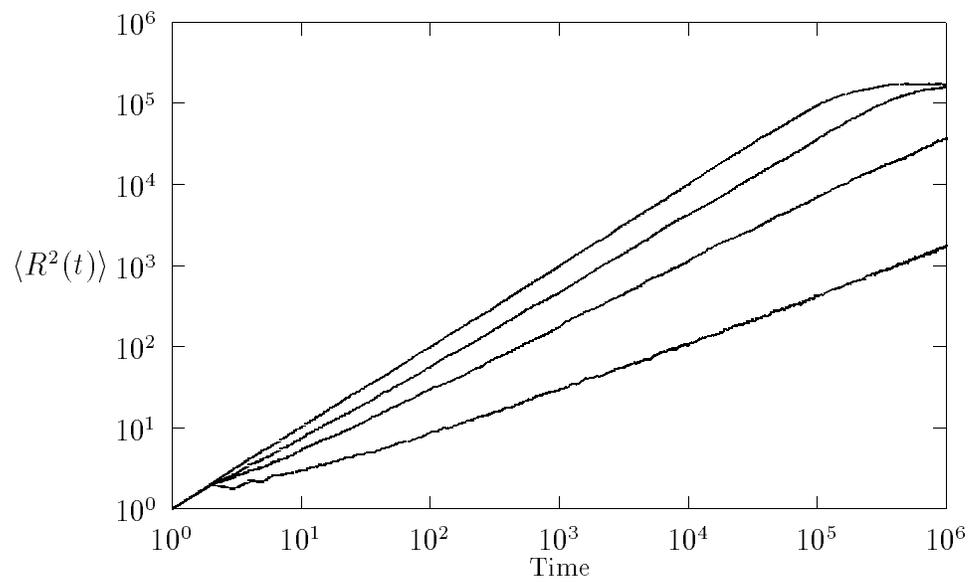
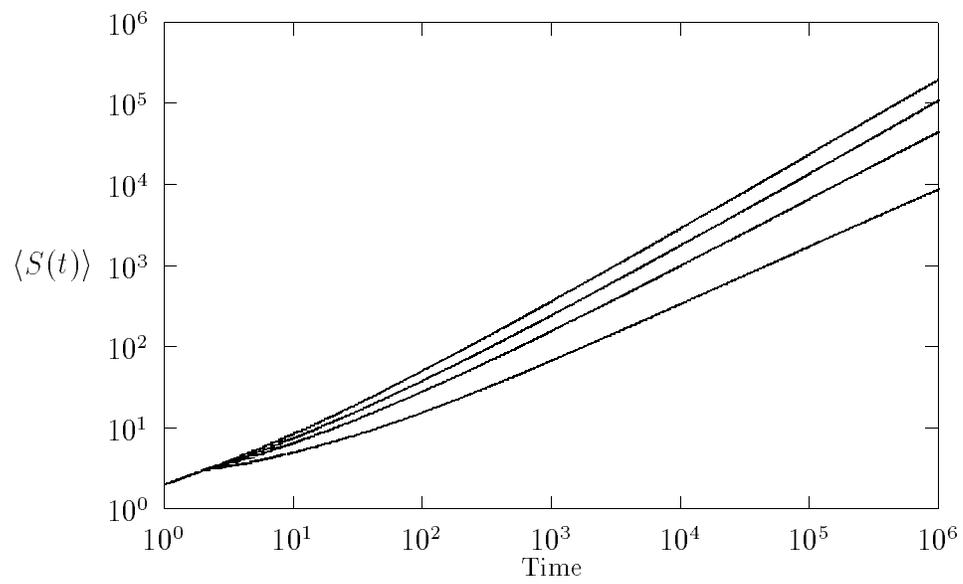
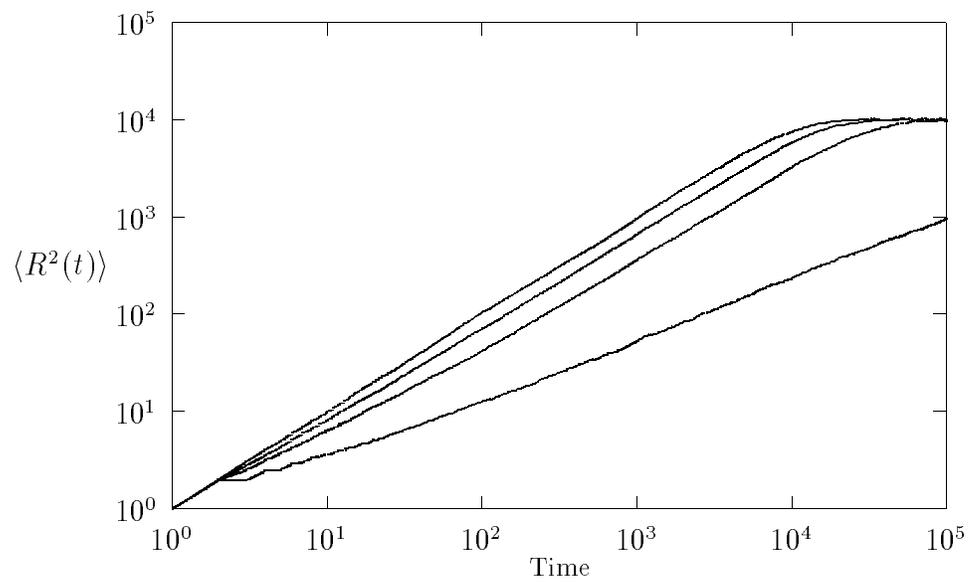
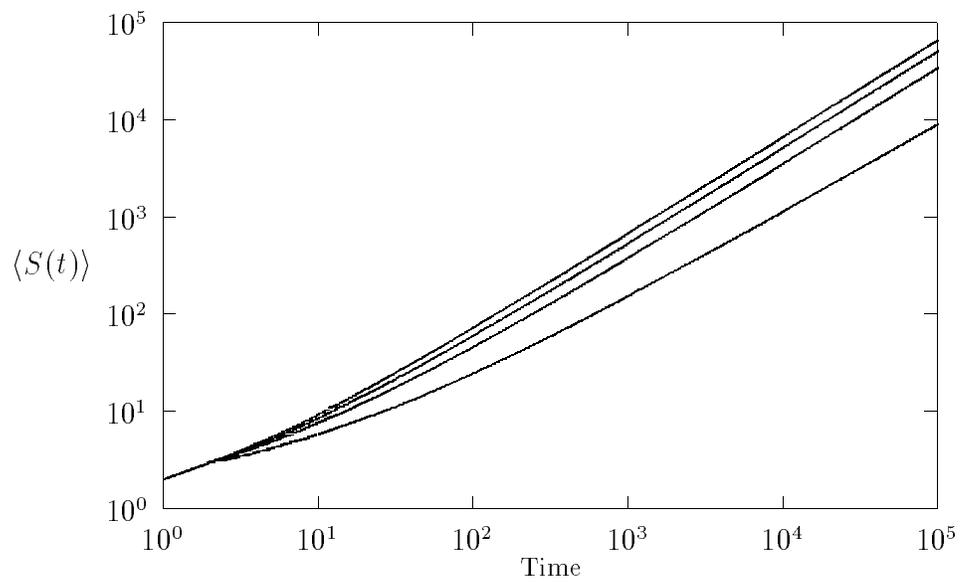
\end{document}